\documentclass[10pt,conference]{IEEEtran} 
\IEEEoverridecommandlockouts

\usepackage{xcolor}
\usepackage{graphicx}
\usepackage{enumitem}
\usepackage{xspace}
\usepackage[T1]{fontenc}
\usepackage{cite}
\usepackage{pifont}
\usepackage{balance}

\newcommand{\tool}{\textsc{DataTrust}\xspace}

\newcommand{\relatedto}{$\,\circ\,$}

\newcommand{\eg}{\textit{e.g.},\xspace}

\newcommand{\etal}{\textit{et al.}\xspace}

\begin{document}

\title{Towards Trustworthy LLMs for Code: \\A Data-Centric Synergistic Auditing Framework}

\author{
\IEEEauthorblockN{Chong Wang$^{1}$, Zhenpeng Chen$^{1}$, Tianlin Li$^{1}$, Yilun Zhang$^{2}$, Yang Liu$^{1}$}
\IEEEauthorblockA{
    $^{1}$\textit{Nanyang Technological University, Singapore}\\
    $^{2}$\textit{AIXpert}\\
    \{chong.wang, zhenpeng.chen, yangliu\}@ntu.edu.sg, tianlin001@e.ntu.edu.sg, yilun@aixpert.io
}
}
\maketitle

\begin{abstract}
LLM-powered coding and development assistants have become prevalent to programmers' workflows. However, concerns about the trustworthiness of LLMs for code persist despite their widespread use. Much of the existing research focused on either training or evaluation, raising questions about whether stakeholders in training and evaluation align in their understanding of model trustworthiness and whether they can move toward a unified direction. In this paper, we propose a vision for a unified trustworthiness auditing framework, \tool, which adopts a data-centric approach that synergistically emphasizes both training and evaluation data and their correlations. \tool aims to connect model trustworthiness indicators in evaluation with data quality indicators in training. It autonomously inspects training data and evaluates model trustworthiness using synthesized data, attributing potential causes from specific evaluation data to corresponding training data and refining indicator connections. Additionally, a trustworthiness arena powered by \tool will engage crowdsourced input and deliver quantitative outcomes. We outline the benefits that various stakeholders can gain from \tool and discuss the challenges and opportunities it presents.
\end{abstract}

\maketitle

\section{Trustworthiness Auditing of LLMs for Code}
Large language models (LLMs) for code~\cite{chen2021evaluating,roziere2023code,guo2024deepseek,zhu2024deepseek,li2023starcoder} have demonstrated significant potential in supporting various stages of the software development lifecycle~\cite{hou2023large,liu2024large,huang2024agents,jin2024mare,ronanki2024requirements,jin2024llms,liu2024your,tamberg2024harnessing,ma2024llmparser,desai2023reinforcement,wang2024teaching,wu2024versicode,zhang2024empirical,wang2023boosting,wang2024rlcoder,zheng2023survey,wang2024tiger,du2024vul}.
As a result, LLM-powered coding and development assistants are now widely integrated into programmers' daily workflows. A prominent example is GitHub Copilot~\cite{copilot}, an LLM-based coding assistant adopted by over 77,000 businesses and downloaded more than 20.3 million times from the VSCode Plugin marketplace (data as of September 24, 2024).

Despite the widespread adoption of LLMs for code in real-world development, significant concerns persist regarding their trustworthiness, particularly in dimensions such as \textit{robustness}, \textit{security}, \textit{timeliness}, \textit{privacy}, \textit{fairness}, etc. In addition to the inherent risks common to general-purpose LLMs (\eg jailbreaking threats)~\cite{wang2023decodingtrust,bommasani2023foundation,zeng2024ai,sun2024trustllm}, LLMs for code introduce additional concerns specific to code and software development. For example, the code generated by these models may contain vulnerabilities or weaknesses that pose serious code security risks~\cite{pearce2022asleep}. Therefore, advancing towards more trustworthy LLMs for code has become an increasingly pressing issue. 

While state-of-the-art research has concentrated on specific trustworthiness dimensions in either \textit{training}~\cite{guo2024deepseek,zhu2024deepseek} or \textit{evaluation}~\cite{huang2024your,wang2024and,pearce2022asleep,yang2024robustness}, the modern development and application of LLMs have evolved into a complex, iterative process of model training (including fine-tuning) and evaluation, involving multiple stakeholders. \textit{This raises crucial questions: Have stakeholders in training and evaluation aligned their understandings of model trustworthiness? How can they be systematically guided toward a unified direction rather than relying on heuristic attempts? To address these, a unified trustworthiness auditing framework is essential—one that synergistically integrates both training and evaluation processes along with their iterative cycles.} Specifically, we present a \textit{data-centric vision} for this framework, named \tool, which attempts to connect the trustworthiness indicators in evaluation to the quality indicators in training. Building on this foundation, \tool autonomously conducts independent inspections of training data and evaluates model trustworthiness using synthesized data, attributing potential causes identified in specific evaluation data to their corresponding training data and aiding in the refinement of indicator connections. A trustworthiness arena powered by \tool can be launched to further engage crowd sources and deliver quantitative and comparative auditing outcomes such as leaderboards.

\begin{figure}
    \centering
    \includegraphics[width=1\columnwidth]{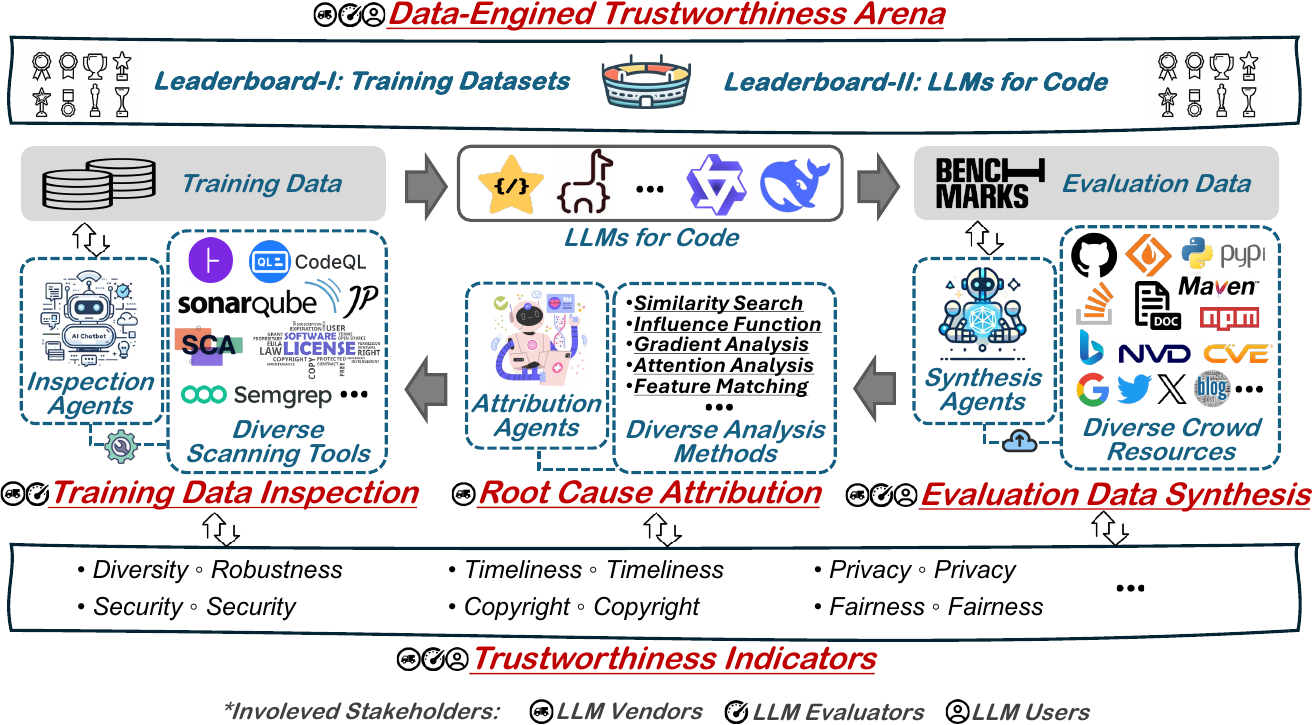}
    \vspace{-5mm}
    \caption{Methodology Overview of \tool.}
    \label{fig:overview}
    \vspace{-5mm}
\end{figure}

Through synergistic auditing, \tool can deliver benefits to a range of stakeholders:
\begin{itemize}[leftmargin=*]
    \item \textit{LLM Vendors.} For vendors controlling both training data and production LLMs, \tool streamlines the development of trustworthy LLMs for code. It provides insights into the cycle of training data inspection, filtering, model training, evaluation, and root cause analysis. Furthermore, publishing auditing reports for both training data and models enhances trustworthiness transparency for downstream stakeholders.
    \item \textit{LLM Evaluators.} \tool offers comprehensive trustworthiness evaluations, supporting continuous trustworthiness report (\eg leaderboards) updates for both commercial and open-source LLMs and training corpora. It provides valuable insights for downstream stakeholders in selecting models or data. Moreover, identified trustworthiness issues can help upstream stakeholders, like LLM vendors, conduct root cause analysis on their training data.
    \item \textit{LLM Users.} \tool engages LLM users in two ways: by enabling them to review auditing reports or leaderboards for insights into model trustworthiness, and by collecting trustworthiness issues encountered by users to create a crowd-sourced trustworthiness arena, which helps evaluators refine assessments and vendors perform root cause analysis.
\end{itemize}

This paper calls for collaboration between academic and industrial communities to refine the data-centric vision and tackle challenges in implementing the \tool framework for trustworthy LLMs for code. The goal is to align understandings, standardize auditing processes, and enhance the transparency of LLM trustworthiness and their data, ultimately benefiting a wide range of stakeholders.

\section{\tool: Methodology and Challenges}
Figure~\ref{fig:overview} presents an overview of \tool's methodology. We start by compiling a comprehensive set of \textit{Model Trustworthiness Indicators} across various dimensions and linking them to corresponding \textit{Training Data Quality Indicators}. These indicators guide (i) the assessment of training data quality and (ii) the construction of thorough evaluation data. Based on this foundation, we design three key data-centric processes: \textit{Training Data Inspection}, \textit{Evaluation Data Synthesis}, and \textit{Root Cause Attribution}. These processes iteratively operate during the training and evaluation phases of LLMs for code, continuously refining indicator connections while engaging multiple stakeholders. Instead of starting from scratch, we will leverage existing scanning tools, crowd-sourced resources, and analysis methods, integrating them with advanced technologies like LLM-based agents. Additionally, we introduce a \textit{Data-Driven Trustworthiness Arena} to engage users actively, enhancing the comparison and benchmarking of mainstream LLMs for code and their training datasets. The following section further explores the methodology and highlights key challenges.


\subsection{Trustworthiness Indicators and Connections}
Each entry is represented as an indicator pair \(\mathcal{T}\)\relatedto\(\mathcal{E}\), where \(\mathcal{T}\) denotes a quality indicator for training data and \(\mathcal{E}\) represents a trustworthiness indicator for model evaluation.

\begin{itemize}[leftmargin=*]  
    \item \textit{Diversity\relatedto Robustness.} Training data often lacks diversity, leading to imbalanced distributions across domains, functionalities, and identifiers. This can raise model robustness concerns, such as inconsistent performance across domains~\cite{zhuo2024bigcodebench} and vulnerabilities to adversarial attacks~\cite{yang2022natural}.
    \item \textit{Security\relatedto Security.} Security risks, including various software vulnerabilities are commonly found in open-source code repositories and may remain undetected for extended periods~\cite{pearce2022asleep}. Consequently, the presence of insecure data samples in the training data might result in LLMs generating insecure outputs, such as vulnerable code. 
    \item \textit{Timeliness\relatedto Timeliness.} Training data sourced from open-source code repositories over extended periods may include outdated information, such as the use of outdated code patterns, inactive libraries, and deprecated APIs~\cite{wang2024and}. This might cause the trained LLMs to generate outdated outputs. 
    \item \textit{Copyright\relatedto Copyright.} Source files in training data may be subject to specific licensing terms or copied from other licensed repositories, raising two major copyright concerns. First, the training data might introduce the risk of license term violations, particularly when used for training commercial LLMs~\cite{rahman2023beyond}. Second, the trained LLMs might generate outputs that raise user concerns regarding copyright~\cite{li2024purifying}.  
    \item \textit{Privacy\relatedto Privacy.} Training data often contains hard-coded privacy-sensitive information, particularly software credentials (\eg API keys and passwords) and personally identifiable information (\eg names and addresses), which LLMs might inadvertently reproduce during inference~\cite{huang2024your}.  
    \item \textit{Fairness\relatedto Fairness.} Machine learning algorithms are frequently reported to exhibit fairness issues related to protected demographic attributes \cite{chen2024fairness}. LLMs for code, which are trained on historical code data, have been shown to perpetuate these biases, generating code/algorithms that may discriminate against certain demographic groups \cite{huang2023bias,liu2023uncovering}.
    
\end{itemize}  

\ding{47} \textbf{\textit{Challenge 1.}} The complex working mechanisms of LLMs make it challenging to construct precise and fine-grained connections between model trustworthiness indicators and training data quality indicators. We have drafted initial relationships based on existing studies~\cite{yang2024robustness}. This challenge also motivates our call to action for the synergistic auditing framework that integrates both training and evaluation processes. Through \tool, we believe clearer, more systematic connections between these indicators can be gradually identified and refined through collaboration between relevant communities. Ultimately, this will significantly enhance the development and refinement of trustworthy LLMs for code.

\subsection{Training Data Inspection}
Based on these indicators, \tool automatically inspects the training data of LLMs for code to identify potential quality issues that could impact the model trustworthiness.

\textbf{Integrating Diverse Scanning Tools.} 
We can integrate a range of scanning and analysis tools to conduct targeted data inspections. These include code element extraction and semantics annotation for profiling data diversity, static vulnerability scanning for assessing security, software composition analysis (SCA) for tracking library and API versions, clone detection for analyzing license violations and establishing traceability, and regular expression generation for detecting privacy-sensitive information, fairness analysis, among others. 

\ding{47} \textbf{\textit{Challenge 2.}} Scalability is a challenge when scanning large training corpora, which can contain millions of files (\eg 603 million in DeepSeek-Coder's corpus~\cite{guo2024deepseek,zhu2024deepseek}), leading to resource and time constraints. A practical solution is to sample a portion of the data for initial inspection, providing an approximate quality assessment.

\textbf{LLM-based Inspection Agents.} 
The results generated by existing scanning tools are often not directly usable as measurable quality indicators and need to be refined and aggregated. LLM-based agents are well-suited to handle this task. For example, an agent that explores the training corpus and understands code semantics through static analysis (\eg identifier extraction) is crucial for profiling data diversity. Another agent is required to count deprecated libraries or APIs by cross-referencing online API documentation, based on library and API version tracing results from SCA tools. 

\ding{47} \textbf{\textit{Challenge 3.}} Existing scanning tools have varied prerequisites, input/output formats, and configuration options, making integration challenging but offering innovation opportunities. For instance, Infer~\cite{infer} requires a compilable code project as input and allows customizable detection settings. To enhance integration, future efforts could focus on unified input/output formats or domain-specific languages (DSLs) for tool and agent communication, on-demand configuration selection via agents for more flexible inspections, and aggregating results from different tools for more reliable decisions.


\subsection{Evaluation Data Synthesis}
To audit the trustworthiness of LLMs for code across various indicators, \tool automatically synthesizes evaluation data and benchmarks by referencing a wide range of evolving resources.

\textbf{Aggregating Diverse Crowd Resources.} In addition to existing evaluations that rely on (semi-)manually curated benchmarks~\cite{zhuo2024bigcodebench,du2024evaluating}, \tool dynamically generates comprehensive, up-to-date evaluation data by leveraging diverse real-time sources. These resources include open-source software platforms, API documentation, package management tools, developer Q\&A forums, vulnerability databases, search engines, and social media. 
By gathering insightful information related to coding and software development, \tool can inspire concrete test cases and deliver a more adaptable trustworthiness auditing. For example, it retrieves domain-specific coding task descriptions from GitHub to evaluate the domain robustness of the LLM under test (LLMUT). For security, recent malware and vulnerability reports are sourced from package managers, vulnerability databases, and social media. 
Additionally, discussions on other concerns across these resources inform evaluations of specific dimensions.

\textbf{LLM-based Synthesis Agents.} Diverse resources must be processed into \textit{executable} test cases, which \tool achieves using LLM-based agents. Each agent handles data synthesis for a specific evaluation dimension, with modules for automated data fetching (\eg vulnerability disclosures), test case generation (\eg probing prompts and oracles), and result inspection. For example, in deprecated API usage evaluation, the agent fetches release notes (\eg from Libraries.io~\cite{libraries.io}) to identify deprecated APIs, retrieves related code snippets from sources like GitHub, and generates prompts to test whether the LLM under test still uses the deprecated APIs.

\ding{47} \textbf{\textit{Challenge 4.}} Crowd resources, like informal text from Stack Overflow, are heterogeneous and challenging to integrate. This opens research opportunities, such as creating unified intermediate representations (IRs), developing adaptive crawling and parsing methods, and refining information fusion and conflict resolution. These tasks could leverage LLMs for web exploration, code generation (\eg HTML parsing code), and information summarization.


\subsection{Root Cause Attribution}
The trustworthiness issues identified during evaluation are often connected to specific instances within the training data. We use root cause attribution to link trustworthiness issues identified during evaluation to problematic instances in the training data. This is crucial for systematic trustworthiness auditing and developing trustworthy LLMs for code because it (i) refines the connections between model trustworthiness and training data quality, (ii) addresses incomplete training data inspections by identifying previously unknown problematic instances, and (iii) provides clearer insights into which training instances significantly impact the model's trustworthiness when combined with direct inspection results.

\textbf{Adopting Diverse Attribution Methods.} Various instance-level attribution methods exist to assess the correlation between evaluation instances and training instances. These methods include influence functions~\cite{koh2017understanding}, similarity search~\cite{pezeshkpour2021empirical}, gradient-based analysis~\cite{pruthi2020estimating}, attention analysis~\cite{hao2021self}, each with its own strengths and weaknesses, and some may overlap in functionality. In addition to instance-level attribution, the identified issues can also facilitate pattern-level attribution, enabling the identification of specific categories of unknown problematic training data. For each identified issues, we analyze the common patterns contributing to the issues and translate them into specific, lightweight inspection rules (\eg CodeQL~\cite{codeql} queries for particular types of vulnerabilities). This also acts as an incremental inspection mechanism, addressing scalability limitations in training data inspection.

\textbf{LLM-based Attribution Agents.} To leverage the advantages of these diverse methods, we employ LLM-based agents to perform comprehensive analyses based on the outputs of these methods, ultimately making informed decisions through the LLM's understanding and summarization capabilities for both code and natural language. For instance, if a code security issue is detected during evaluation (such as the recurrence of a newly disclosed vulnerability in LLMUT-generated code), an attribution agent for code security first utilizes existing instance-level attribution methods to identify training instances (such as code snippets or functions) that may have contributed to the issue. Subsequently, the agent determines the final instances related to the issue by comparing the code patterns with the identified vulnerability patterns, leveraging its code understanding capabilities.

\ding{47} \textbf{\textit{Challenge 5.}} Although instance-level attribution methods like influence functions and gradient-based analysis have shown promise in other domains~\cite{dai2023training,pezeshkpour2021empirical}, their performance in code-related data analysis remains untested. This creates risks when applying them for root cause attribution in LLMs for code, but also presents opportunities for refining these methods to suit code-specific challenges. Additionally, effectively combining instance-level analysis with pattern-level attribution for incremental training data inspection is a challenge. A mechanism is needed to ensure the accuracy of patterns extracted from evaluation issues, guiding more reliable pattern-level attribution.

\subsection{Data-Engined Trustworthiness Arena}
Based on the framework outlined above, \tool can also provide a data-driven trustworthiness arena inspired by the Chatbot Arena~\cite{chiang2024chatbot}. This arena features two leaderboards: one for LLMs for code and another for training data. It continuously maintains and updates these leaderboards by applying the three data-centric processes to various LLMs for code and their corresponding training datasets when applicable. For LLMs with non-open-sourced training data, \tool can still rank them on the LLMs for code leaderboard using evaluation data synthesis alone.

\textbf{Engaging Crowd User Interactions.} Additionally, \tool provides interfaces for community contributions like Chatbot Arena, related to test case creation, oracle validation, and result confirmation. The key to this initiative is designing suitable interaction paradigms—such as engaging mini-games or daily coding tasks—to minimize participant difficulties and reduce manual efforts for users. To facilitate this, LLM-based agents are employed to offer guidance and assistance for crafting and manipulating test inputs, as well as to automatically convert human-involved tasks into user-friendly or user-transparently information formats. For instance, if the data synthesis agent for code security generates a test input and identifies a recurring vulnerability in the code produced by the LLMUT, an assistance agent in the arena can gather additional relevant information about this vulnerability from diverse resources and present it as a concise, readable checklist for community contributors.

\ding{47} \textbf{\textit{Challenge 6.}} Unlike the Chatbot Arena, where tasks are simple for general users, our trustworthiness arena faces a higher participation threshold due to the complexity of certain dimensions like code security. Even experienced developers may struggle to identify or verify vulnerabilities in code. To address this, we need to design more accessible interaction paradigm by simplifying or restructuring the tasks. For example, instead of relying solely on yes/no binary annotations for potentially vulnerable code, we can provide a checklist of low-level safeguard operations (\eg index range validation) and ask users to verify each item.

\section{Related Works}
Trustworthiness issues in general LLMs have garnered significant attention from both academia and industry. Existing research has introduced various principles and dimensions of trustworthiness and benchmarked several mainstream LLMs~\cite{sun2024trustllm,hong2024decoding,wang2023decodingtrust}. These dimensions often include \textit{safety/security}, \textit{fairness}, \textit{robustness}, \textit{privacy}, \textit{machine ethics}, \textit{transparency}, \textit{accountability}, and \textit{regulations and laws}. Since LLMs for code are frequently built by fine-tuning general LLMs or by training on both text and code corpora, they inherently inherit these broader trustworthiness concerns. For instance, LLMs for code may be vulnerable to adversarial attacks, which can undermine their robustness~\cite{yang2022natural}.

Within the software engineering community, trustworthiness in LLMs for code is also becoming an increasingly important issue. Lo~\cite{lo2023trustworthy} has called for trustworthy and synergistic AI for Software Engineering (AI4SE), offering a systematic overview of many open challenges and opportunities. Yang \etal~\cite{yang2024robustness} revisit the dimensions of trustworthiness in LLMs for code, covering aspects such as \textit{robustness}, \textit{security}, \textit{privacy}, \textit{explainability}, \textit{efficiency}, and \textit{usability}, and suggest initial enhancement principles focusing on training data. Spiess \etal~\cite{spiess2024calibration} have introduced correctness calibration techniques for LLMs for code to improve the trustworthiness of their outputs. Additionally, other studies address specific trustworthiness dimensions and application areas~\cite{zhao2024models}, such as trustworthy program synthesis~\cite{key2022toward}, backdoor-trigger taxonomy~\cite{hussain2024trojans}, and trustworthy code summarization~\cite{virk2024enhancing}.

These works provide a strong foundation for realizing our unified auditing framework, \tool. Our primary objective is to align understanding, standardize auditing processes, and enhance the transparency of the trustworthiness of LLMs and their data, ultimately benefiting a broad spectrum of stakeholders.

\section{Summary}
In this paper, we propose a vision for a unified trustworthiness auditing framework, \tool, which adopts a data-centric approach that synergistically emphasizes the relationship between training and evaluation data. \tool seeks to connect model trustworthiness indicators in evaluation with data quality indicators in training. It autonomously inspects training data, evaluates model trustworthiness using synthesized data, and attributes potential causes from specific evaluation data to their corresponding training data while refining indicator connections. \tool can achieve \textit{extensibility} through the integration of diverse tools and resources, as well as \textit{evolvability} by incorporating real-world information and knowledge. Nevertheless, open challenges remain, particularly concerning tool integration, resource aggregation, method adoption, and interaction paradigms, all of which present valuable opportunities for future research. We believe this vision and \tool can deliver benefits to a broad array of stakeholders involved in both the development and application of trustworthy LLMs for code.
\section{Future Plans}
We propose several actionable plans with achievable timelines. First, we will focus on a subset of \tool to initiate exploration and implementation, starting with a prototype. This subset will include a few key indicators, such as security and copyright, applied to open-source LLMs for code (\eg StarCoder) alongside open-source training data, and a well-established application task such as code generation. Second, we will implement the three data-centric auditing processes, using the selected indicators, LLMs, and application task, designing approaches to address the challenges identified earlier. Third, we will launch an initial arena platform to engage participants, providing continuously updated leaderboards. Finally, we will refine the methodology and complete \tool through collaboration and feedback from the broader community.

\bibliographystyle{IEEEtran}
\bibliography{ref}
\balance
\end{document}